# Comment on "Experimental motivation and empirical consistency in minimal no-collapse quantum mechanics"

Fred H. Thaheld
99 Cable Circle, #20
Folsom, Calif. 95630 USA
fthaheld@directcon.net

**Abstract**

Schlosshauer has advanced a theory of minimal no-collapse quantum mechanics for a decoherence-based subjective resolution of the "measurement problem". The basic premise being that superposition states are maintained beyond the retinal apparatus, becoming correlated with neuronal arrays located in the occipital lobe of the brain. Decoherence for these neurons in a superposition of firing and resting, leads to an irreversible dynamical decoupling of the two branches resulting in the emergence of a single subjective perception. Based upon prior retinal research, it is shown that his theory is untenable for several reasons.

In a recent paper Schlosshauer (1) has proposed that the perception of single "outcomes" resulting from superposed states, can be explained through decoherence effects in the neuronal apparatus in the occipital cortex of the brain, bypassing the need for a collapse mechanism. In a footnote at the bottom of p.20, he referenced a paper of mine (2), among several other authors, which reads as follows: "A search for experimental evidence for such superpositions has been suggested in [146]; for an experimental proposal, see [147]. Cf. also [148] (Thaheld) for an (unconvincing) suggestion that the visual apparatus itself might trigger a physical collapse".

The ironic thing about this footnote is that a prior reference which he cited [146], is a paper written by Abner Shimony (3), who theorized that the "locus of reduction (collapse) is the macromolecules of the sensory and cognitive facilities". Shimony concludes by saying that, "My conjecture is that the reduction occurs at the retinal molecule itself: that there is a superselection rule operative which prevents a superposition of molecular conformations as different as *cis* and *trans* from occurring in nature". In my paper I advanced exactly the same theory (2) with additional supporting references and analysis, along with a proposed experiment to test for same, while at the same time providing attribution for Shimony's prescient observation. One might now assume that both Shimony and I have made an unconvincing suggestion regarding a possible collapse option.

Schlosshauer appears to be incorrect in several statements he has made in advancing his theory of no-collapse quantum mechanics for a decoherence-based subjective resolution of the "measurement problem." In Sec. IV C, Schematic sketch of the chain of interactions in visual perception and cognition, he states that "the photon-rhodopsin interaction should lead to an (albeit, due to the influence of decoherence, very fragile) superposition of the different biochemically distinct states of the rhodopsin molecule. These relative states can then be expected to be further correlated with the appropriate states of neuronal arrays that are mainly located in the primary visual area in the occipital lobe of the brain". This is practically the equivalent of Ghirardi's proposal (4) that superposed states would somehow persist beyond the retina into the visual cortex, where Spontaneous Localization would result in a collapse. I have previously outlined how his theory was incorrect, and how one might perform an experiment in an attempt to resolve this issue (5).

Schlosshauer's approach differs in that he theorizes that the extreme fast decoherence rate for certain neurons in the brain, in a superposition of firing and resting, will lead to a practically irreversible dynamical decoupling of the two branches that now describe two distinct "outcomes". This not only runs counter to what Shimony and I have conjectured but, to extensive retinal research conducted over a period of several decades.

He has stated that, "Due to their mesoscopic properties, rhodopsin molecules are subject to strong decoherence, such that already at this stage the influence of the environment will have preselected the robust states of the rhodopsin molecule, which correspond to certain photon detection events". This is incorrect in that the complete transition of the rhodopsin molecule from *cis* to *trans*, from microscopic to mesoscopic to macroscopic takes place in about 200 fs (6). The molecular shape change is the classical signal of the photoabsorption quantum event. In addition, the vibrational spectrum also evolves as the molecule is changing its shape. The photon interaction that he proposes is not of a superposed state, as there would have already been a collapse of the wave function of the superposed photon state prior to this point in time. Even if one is generous and accepts his theory up to this point, there are several additional reasons why a superposed state could never get as far as the occipital cortex (2,5).

First, the response of the retinal rods to a single photon is initiated by activation of a single rhodopsin molecule, and results in an amplified electrical response about 1 pA in amplitude and about 1 s long. One activated rhodopsin molecule causes approximately $10^7$ charges (sodium ions) to fail to enter the cell. This response is what causes anywhere from 2 to 3 action potentials, or retinal ganglion cell (RGC) spike trains, to be generated by the ganglion cell, where they then proceed to the optic nerve (7). This alone would pose an immediate hazard for any superposed state, from the standpoint of trying to avoid immediate decoherence of the superposed branches, either as a result of the massive number of sodium ions or, depending upon whether one gets 2 or 3 action potentials each time.

Second, the retina contains 5 major types of neurons (covered in more detail in (5,8,9), consisting of rod and cone photoreceptors connected to bipolar as well as horizontal cells, with the bipolar cells in turn making synapses with amacrine cells and retinal ganglion cells, whose axons form the optic nerve. Photoreceptors send electrochemical signals to the brain by both cone and rod synaptic mechanisms. There also exist very sophisticated modulation systems that are facilitated by the horizontal and amacrine cells. As a further example of this complexity, the dendrites of ganglion cells reach up into the inner retina and read out activity formed by the interactions of bipolar and amacrine cells. What this means in effect, is that it appears that it would be very difficult, if not impossible, to maintain any superposed states, whether they would consist of nervous signals or of so called "biochemically distinct states of the rhodopsin molecule", so that they can end up in the occipital cortex of the brain, if they are faced with passage through this multitude of barriers.

Third, the information in the RGC spike trains is relayed to the visual cortex by lateral geniculate nucleus (LGN) relay cells operating in either of two regimes: tonic mode where each RGC spike is relayed by a single LGN spike or, burst mode, where a single RGC input spike is relayed as a stereotyped burst of spikes (2,10). It is highly doubtful as to whether any superposed state, provided it even got this far, could survive being converted into either a tonic or burst mode, which represents quite a drastic transition.

In conclusion, I would like to make the following observation. It is extremely difficult to attempt to extrapolate any results obtained in the much simpler, inanimate experimental domains of SQUIDs, molecular interferometry and Bose-Einstein condensation, to the exceedingly more complex animate domain of the brain. That is why I have designed an experiment which I hope will finally allow us to resolve this conflict between the adherents of either a collapse or decoherence mechanism (5).

**References**

(1) Schlosshauer, M., 2006. Experimental motivation and empirical consistency in minimal no-collapse quantum mechanics. Ann. Phys. 321, 112-149. quant-ph/0506199 v3.